\documentclass[fleqn,usenatbib]{mnras}

\usepackage{newtxtext,newtxmath}
\usepackage{dblfloatfix}

\usepackage[T1]{fontenc}

\DeclareRobustCommand{\VAN}[3]{#2}
\let\VANthebibliography\thebibliography
\def\thebibliography{\DeclareRobustCommand{\VAN}[3]{##3}\VANthebibliography}

\usepackage{graphicx}	
\usepackage{amsmath}	
\usepackage{multicol}   

\newcommand{\Msun}{\,M$_{\odot}$}

\defcitealias{Ebrova2017}{EL17}
\defcitealias{Revaz2018a}{RJ18}

\title[Origin of stellar prolate rotation]{Origin of stellar prolate rotation in a cosmologically  simulated faint dwarf galaxy}

\author[Salvador Cardona-Barrero et al.]{Salvador Cardona-Barrero,$^{1,2}$\thanks{E-mail: scardona@iac.es}
Giuseppina Battaglia,$^{1,2}$
Arianna Di Cintio,$^{2,1}$\thanks{Junior Leader Fellow ‘La Caixa’ Foundation}
Yves Revaz$^{3}$ \& \newauthor Pascale Jablonka$^{3,4}$
\\
$^{1}$Instituto de Astrof\'isica de Canarias, Calle Via L\'actea s/n, E-38206 La Laguna, Tenerife, Spain\\
$^{2}$Universidad de La Laguna. Avda. Astrof\'isico Fco. S\'anchez, La Laguna, Tenerife, Spain\\
$^{3}$Institute of Physics, Laboratoire d'astrophysique, \'Ecole Polytechnique F\'ed\'erale de Lausanne (EPFL), Observatoire, CH-1290 Versoix, Switzerland \\
$^{4}$GEPI, Observatoire de Paris, Université PSL, CNRS, Place Jules Janssen, F-92190 Meudon, France
}

\date{Accepted XXX. Received YYY; in original form ZZZ}

\pubyear{2020}

\begin{document}
\label{firstpage}
\pagerange{\pageref{firstpage}--\pageref{lastpage}}
\maketitle

\begin{abstract}
Stellar prolate rotation in dwarf galaxies is rather uncommon, with only two known galaxies in the Local Group showing such feature (Phoenix and And~II). 
Cosmological simulations show that in massive early-type galaxies prolate rotation likely arises from major mergers. However, the origin of such kinematics in the dwarf galaxies regime has only been explored using idealized simulations. Here we made use of hydrodynamical cosmological  simulations of dwarfs  galaxies with stellar mass between  $3\times10^5$ and $5\times10^8$~\Msun\ to explore the formation of prolate rotators.
Out of $27$ dwarfs, only one system showed clear rotation around the major axis, whose culprit is a major merger at $z=1.64$, which caused the transition from an oblate to a prolate configuration.
Interestingly, this galaxy displays a steep metallicity gradient, reminiscent of the one measured in Phoenix and And~II: this is the outcome  of the merger event that dynamically heats old, metal-poor stars, and of the centrally concentrated residual star formation. 
Major mergers in dwarf galaxies offer a viable explanation for the formation of such peculiar systems, characterized by steep metallicity gradients and prolate rotation.
\end{abstract}




\section{Introduction}

Kinematic anomalies, such as misaligned rotation axes, counter-rotating cores and sub-structures, can be tell-tales of the occurrence of important, violent events in the life of a galaxy, e.g. mergers \citep[e.g.][]{Cox2006, Shapiro2008, Barrera-Ballesteros2015, Ebrova2020, Nevin2021}.

A characteristics of special interest is the presence of stellar prolate rotation, i.e. when the angular momentum of the stellar component is aligned with its major-axis. 
Prolate rotation seems to be fairly common in massive early type galaxies (ETGs), with its frequency increasing with stellar mass, as seen in observations  \citep[see e.g.][and references there in]{Tsatsi2017,Krajnovic2018} and cosmological simulations such as \texttt{ILLUSTRIS} and \texttt{Magneticum Pathfinder} (see \citealt{Ebrova2017}, here after \citetalias{Ebrova2017}; \citealt{Schulze2018}).
 Both simulations and observations indicate mergers as the culprit for stellar prolate rotation. Theoretical works suggest that, while not all mergers do lead to prolate rotation, its appearance strongly correlates with the time of the last major merger \citepalias{Ebrova2017}, and is more likely to be found in systems that undergo a larger number of gas-poor mergers \citep{Lagos2020}. In the observational counterpart, \cite{Ebrova2021} found signs of recent interactions (i.e. tidal tails, shells, asymmetric stellar halos, etc.) in all  prolate ETGs analyzed in their study.

Even though in a $\Lambda$CMD context merger events are expected to occur fairly frequently even at low galaxy masses \citep{Deason2014,Rodrigez-Gomez2015,Wetzel2015,Martin2021}, so far, a direct detection of such events on the scale of dwarf galaxies is limited to LMC-mass objects \citep{Martinez-Delgado2012,Privon2017,Annibali2019,Carlin2019}. Therefore, it is particularly intriguing to understand whether indeed the presence of features such as prolate rotation can be interpreted as an indirect but secure sign of the occurrence of past mergers.

Among the population of Local Group (LG) dwarf galaxies, only two systems exhibit prolate rotation: a satellite of M31, And~II \citep{Ho2012}, and the Phoenix transition type dwarf \citep{Kacharov2017}, found at $400$~kpc from the Milky Way and likely on its first approach. In both cases, mergers have been invoked to explain such an anomalous rotation, also given the presence of other curious characteristics, such as a stellar stream or  misaligned rotating components in And~II  \citep{Amorisco2014,delPino2017} and the presence of a young stellar component tilted of $90$~deg with respect to the main body in Phoenix \citep{Hidalgo2009, Battaglia2012}.

To the best of our knowledge, in the mass range of LG dwarf galaxies, only idealized simulations have been used to explore the occurrence of prolate rotation. \cite{Lokas2014} used idealized N-body simulations and proposed an evolutionary model for And~II that involved the merger between two equal-mass disky dwarf galaxies, placed in a radial orbit towards each other and with a specific inclination of the discs angular momenta. \cite{Ebrova2015} expanded this analysis to five collisionless simulations, exploring also non-radial orbits and three inclinations, showing that prolate rotation does not need radial orbits to arise and can also be produced in mergers with different relative orientation of the discs. However, equal-mass mergers between two dwarf galaxies are probably not a common occurrence (see, for example, recent simulation results by \citealt{DiCintio2021}) and in general observations suggest that dwarf galaxies at those stellar masses are typically not highly rotating \citep{Wheeler2017, Kirby2019}. 

In this work, we perform a search for stellar prolate rotation in a set of simulated dwarf galaxies of stellar masses akin to those of And~II and Phoenix, using the hydrodynamical cosmological simulations from \cite{Revaz2018a}, here after \citetalias{Revaz2018a}. Our aim is not to specifically reproduce the characteristics of either system, but rather to understand whether prolate rotation can naturally occur  under more realistic configurations for the progenitors, both in terms of their structure/internal kinematics and orbital configuration, i.e. by using fully cosmological initial conditions rather than idealized ones.  In Sect.~$2$ we discuss the simulations analyzed; in Sect.~$3$ we perform the search for prolate rotation, assess its significance and explain its origin and in Sect.~$4$ we look at the effect this merger had in setting the system's metallicity gradient. We present a summary in Sect.~$5$.

\section{Simulations}

We analyzed  $27$ dwarf galaxies from the  high-resolution, zoom-in cosmological  hydrodynamical simulations presented in  \citetalias{Revaz2018a}. The spatial resolution of these simulations has been set to properly resolve dwarf spheroidal galaxies, with the softening length for the gas and dark matter particles being $10$ and $50$~pc~h$^{-1}$, respectively. The simulated galaxies span more than $3$~dex in stellar mass from $3\times10^5$ to $5\times10^8$~\Msun, well representing LG classical dwarfs, of which they have been shown to correctly reproduce many global relations linking galaxy mass, luminosity and mean metallicity as well as the detailed chemical patterns derived from the stellar abundance ratios \citepalias{Revaz2018a}. The simulations have been run using the code \texttt{GEAR} \citep[][]{Revaz2012}, a fully parallel code based on \texttt{Gadget-2} \citep[][]{Springel2005}.  Initial conditions assume a Planck cosmology \citep{Planck2016}. Gas cooling is done through the \texttt{Grackle} cooling library \citep{Smith2017}, while gas heating is included via a redshift-dependent UV background and supernovae thermal feedback. Gas, stellar and dark matter particles have an initial mass of $4096$, $1024$ and $22462$~\Msun~h$^{-1}$, respectively.  Each star particle is formed assuming a \cite{Kroupa2001} initial mass function, and is stochastically sampled using the random discrete scheme described in \cite{Revaz2016}, in order to set the number and time delay of the Type~Ia and  Type~II SNe. These SNe inject 10\% of their energy into the interstellar medium via  thermal blast-wave feedback \citep{Stinson2006}, while  also enriching  neighbour gas particles. Chemical mixing is allowed by employing a  Smooth Metallicity Scheme \citep{Wiersma2009}.

For a detailed description of the characteristics of the runs we refer the reader to \citet{Revaz2012, Revaz2016} and \citetalias{Revaz2018a}.

\begin{table}
    \centering
    \begin{tabular}{l|r@{}lr@{}lr@{}l}
    \hline
         &   Host\hspace{.5em} &$z=0$ &  Host\hspace{.5em} &$z=1.78$ & Satellite\hspace{.5em} &$z=1.78$ \\
         \hline\hline
        M$_{\rm vir}/$\Msun  & $2.8$&$\times10^9$   & $1.6$&$\times10^9$ & $3.4$&$\times10^8$ \\
        M$_{\ast}/$\Msun     & $1.4$&$\times10^7$   & $6.5$&$\times10^6$ & $2.4$&$\times10^5$ \\
        M$_{\rm gas}/$\Msun  & $2.3$&$\times10^7$   & $4.0$&$\times10^7$ & $9.1$&$\times10^4$ \\
        $\left(L_{\rm Major}/L_{\rm Tot}\right)^2$  & $0.$&$780$ & $0.$&$040$ & $0.$&$004$\\
        V$_{\rm rot}^{\rm max}/\left<\sigma_{\rm 3D}\right>$  & $0.$&$19$  & $0.$&$30$ & $0.$&$33$ \\
        b/a       & $0.$&$76$ & $0.$&$64$ & $0.$&$68$ \\
        c/a       & $0.$&$70$ & $0.$&$49$ & $0.$&$51$ \\
        r$_{1/2}/{\rm kpc}$ & $0.$&$70$ & $0.$&$64$ & $0.$&$62$ \\
        R$_{1/2}/{\rm kpc}$ & $0.$&$57$ & $0.$&$48$ & $0.$&$42$ \\ 
        \hline
    \end{tabular}
    \caption{Properties of \texttt{h048} at z=0 (left) and at $z=1.78$ (middle), corresponding to its main progenitor just before the merger analyzed; properties of the accreting  satellite at $z=1.78$ (right). We include the virial mass $M_{\rm vir}$, the stellar mass $M_{\ast}$, the total gas mass $M_{\rm gas}$, the ratio between the angular momentum of the stellar component parallel to the major axis and the total angular momentum $\left(L_{\rm Major}/L_{\rm Tot}\right)$, the rotation velocity at $2$ kpc divided by the $3D$ velocity dispersion $\sigma_{\rm 3D}^2 = \sum_{i}^{3}\sigma_{i}^{2}/3$, the axis ratios $b/a$ and $c/a$, and the $3$D and projected stellar 
    half mass radii r$_{1/2}$ and R$_{1/2}$.}
    \label{tab:dat}
\end{table}

\section{Prolate rotation}

\subsection{Identification}

We have systematically analyzed the set of $27$ simulated dwarf galaxies described above, searching for prolate rotation in their stellar components at all redshifts. 
In order to identify the prolate nature of a simulated galaxy, we have compared the angle ($\theta$) between the stellar angular momentum vector and the orientation of the principal axes of the ellipsoid that best represents the 3D stellar distribution of a specific galaxy. By definition, in a prolate rotating galaxy the major axis and the angular momentum should be parallel (i.e. $\cos\theta\sim1$). We consider a system to be  clearly prolate rotating when $\cos\theta > 0.9$ for more than $2$ consecutive Gyrs. 

In order to compute the principal axes of the stellar component we have implemented an iterative method that computes the inertia tensor of the stellar distribution in a pre-defined spherical aperture of $4$~kpc from the centre of the galaxy\footnote{We have tested different apertures and the results are consistent.}. The eigenvalues of this matrix can then be easily related to the principal axes of an ellipsoid.

With this method, we identified only one system that shows clear signs of prolate rotation. We refer to it as \texttt{h048}, where the  ID is the same as in \citetalias{Revaz2018a}. At $z=0$, this galaxy presents a stellar mass of $1.37\times10^{7}$\Msun\, and a $3$D stellar half-mass radius of $0.70$~kpc. The $3$D average rotation curve, up to $4$~kpc, increases linearly with radius, with the average slope of the rotational velocity being $dv_{\rm rot}/dr =0.88\pm0.12$~km~s$^{-1}$~kpc$^{-1}$,  where $r$ indicates the distance to the major axis. Tab.~\ref{tab:dat}\, summarizes the  main characteristics of this system at different redshifts.

\begin{figure}
    \centering
    \includegraphics[width=.85\columnwidth]{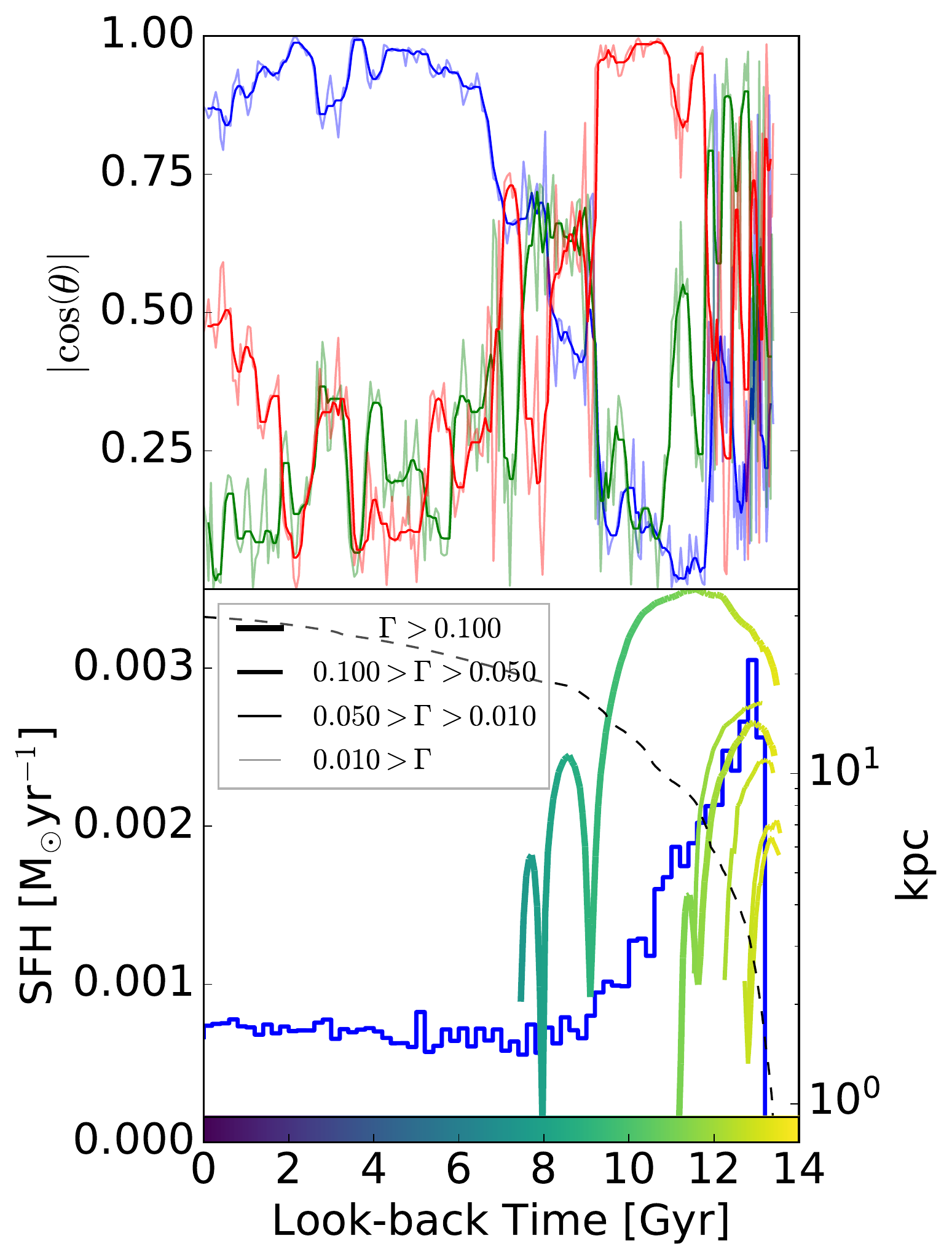}
    \caption{Top: Principal Axes orientation with respect to the stellar angular momentum as a function of look-back time. The major, minor and intermediate axis are shown in blue, red and green respectively. We show the values obtained from individual snapshots (thin lines) and a running average (thick lines) using a window of $6$ snapshots ($\sim0.34$ Gyrs). Bottom: SFH (blue) and dark matter halo virial radius (black dashed line) of \texttt{h048}, together with the orbits of infalling satellites, color-coded by look-back time (the width of the lines has been set depending on the total mass ratio $\Gamma$ between the main progenitor and the satellite,  measured at the snapshot when the satellite first crosses the main progenitor's virial radius). The SFH is similar to the one found in Phoenix \citep{Gallart2015}. Note that the gas mass of the satellite is negligible with respect to that of the host (see Tab.~\ref{tab:dat}), being the effect on the SFH imperceptible. 
    }
    \label{fig:PA}
\end{figure}

Fig.~\ref{fig:PA}, top panel, shows the principal axes orientation with respect to the stellar angular momentum as a function of look-back time. 
It is evident that galaxy \texttt{h048} suffers a transformation at $\sim$10 Gyrs, that completely changes the orientation of its stellar angular momentum vector, from being perpendicular to the major axis ($\cos\theta\sim0$) to being parallel ($\cos\theta\sim1$). The transformation appears to become complete at a look-back time of $\sim6.5$ Gyrs, and remains stable afterwards until $z=0$.

We have also checked that our galaxy fullfills the selection criteria used by 
\citetalias{Ebrova2017}, i.e. that the fraction of angular momentum around the major axis, $L_{\rm Major}$ is larger than $0.5$ of the total angular momentum $L_{T}$ in different apertures. These predefined apertures are $0.25$, $0.35$, $0.5$, $0.7$ and $1r_{\rm max}$, where $r_{\rm max}$ is the radius in which the density is equal to $1\times10^4$~\Msun~kpc$^{-3}$. For our galaxy this value is $r_{\rm max} = 3.9$~kpc. This criterion ensures that the stars rotate around the major axis at all radii not only in the outskirts or in the central regions.


\begin{figure}
    \centering
    \includegraphics[width=0.15\textwidth]{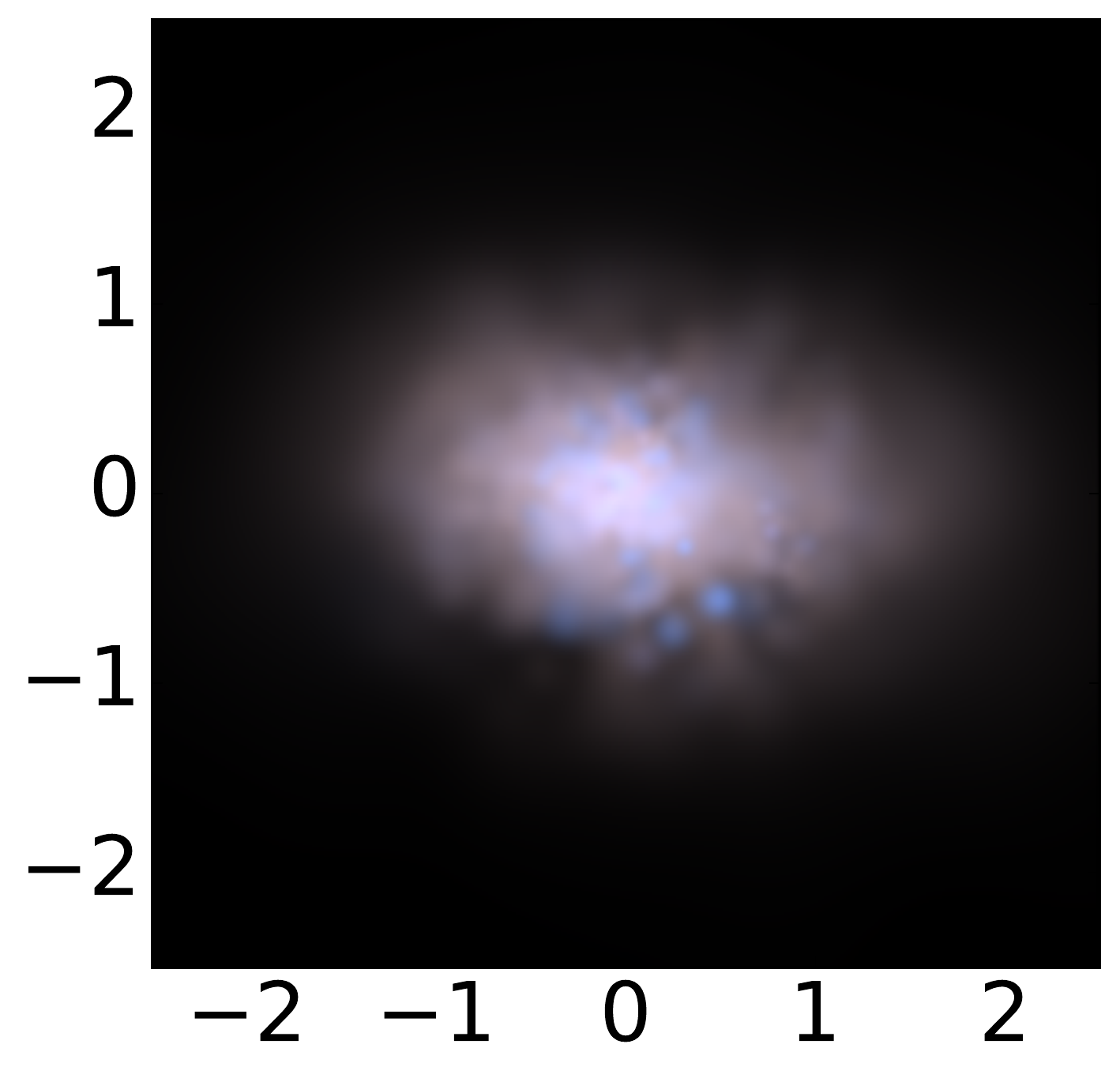}%
    \includegraphics[width=0.15\textwidth]{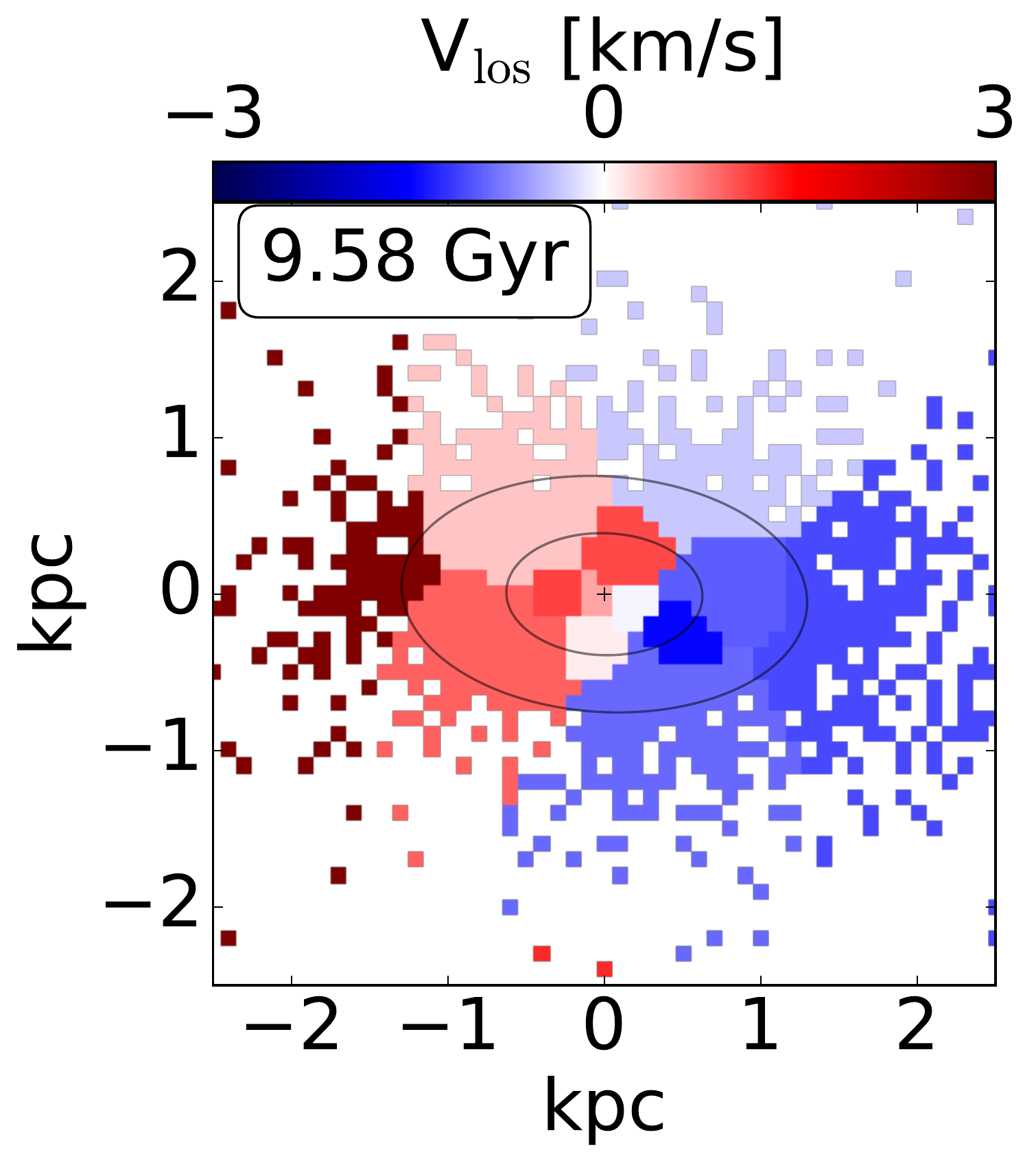}%
    \includegraphics[width=0.15\textwidth]{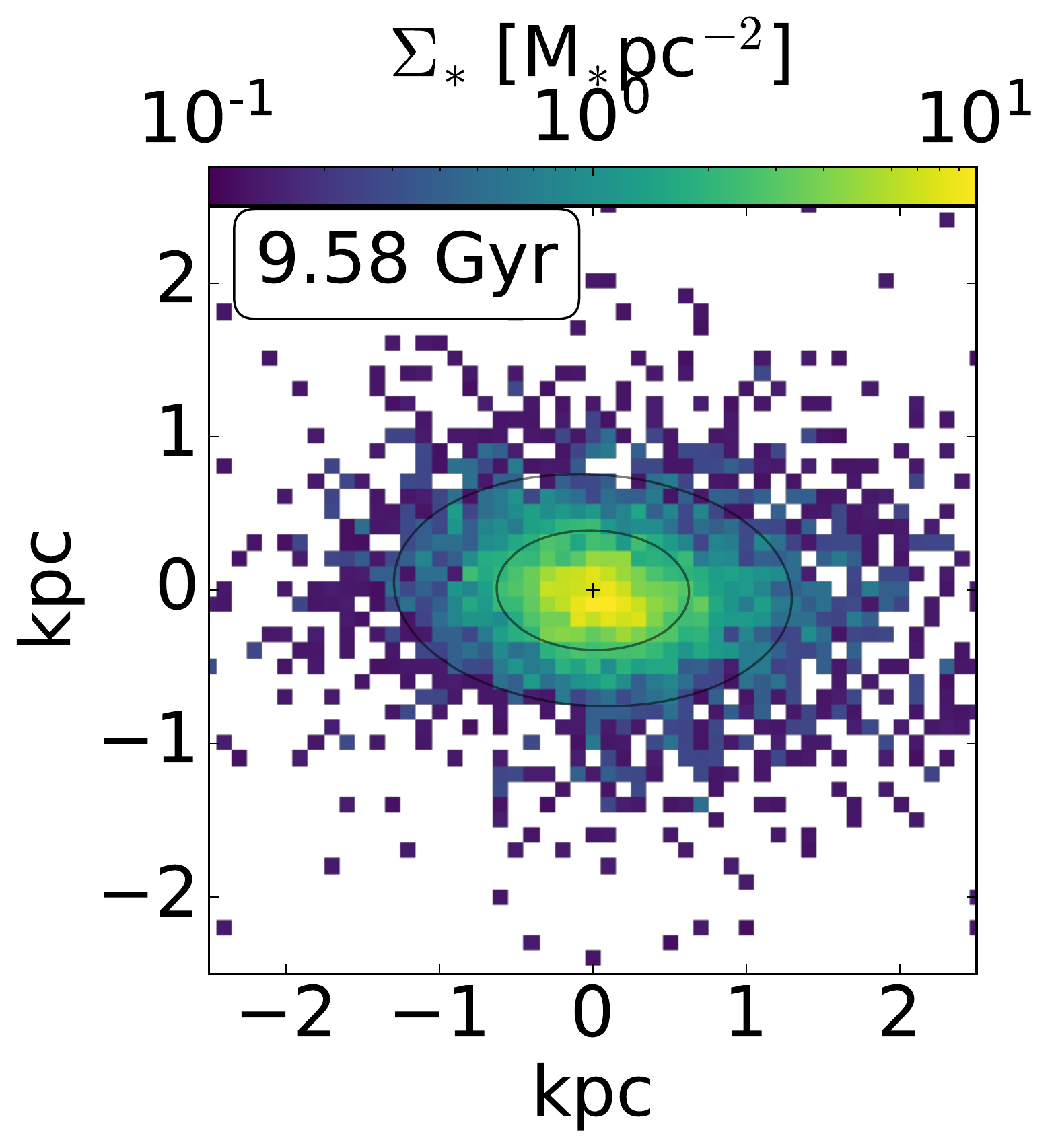}
    \includegraphics[width=0.15\textwidth]{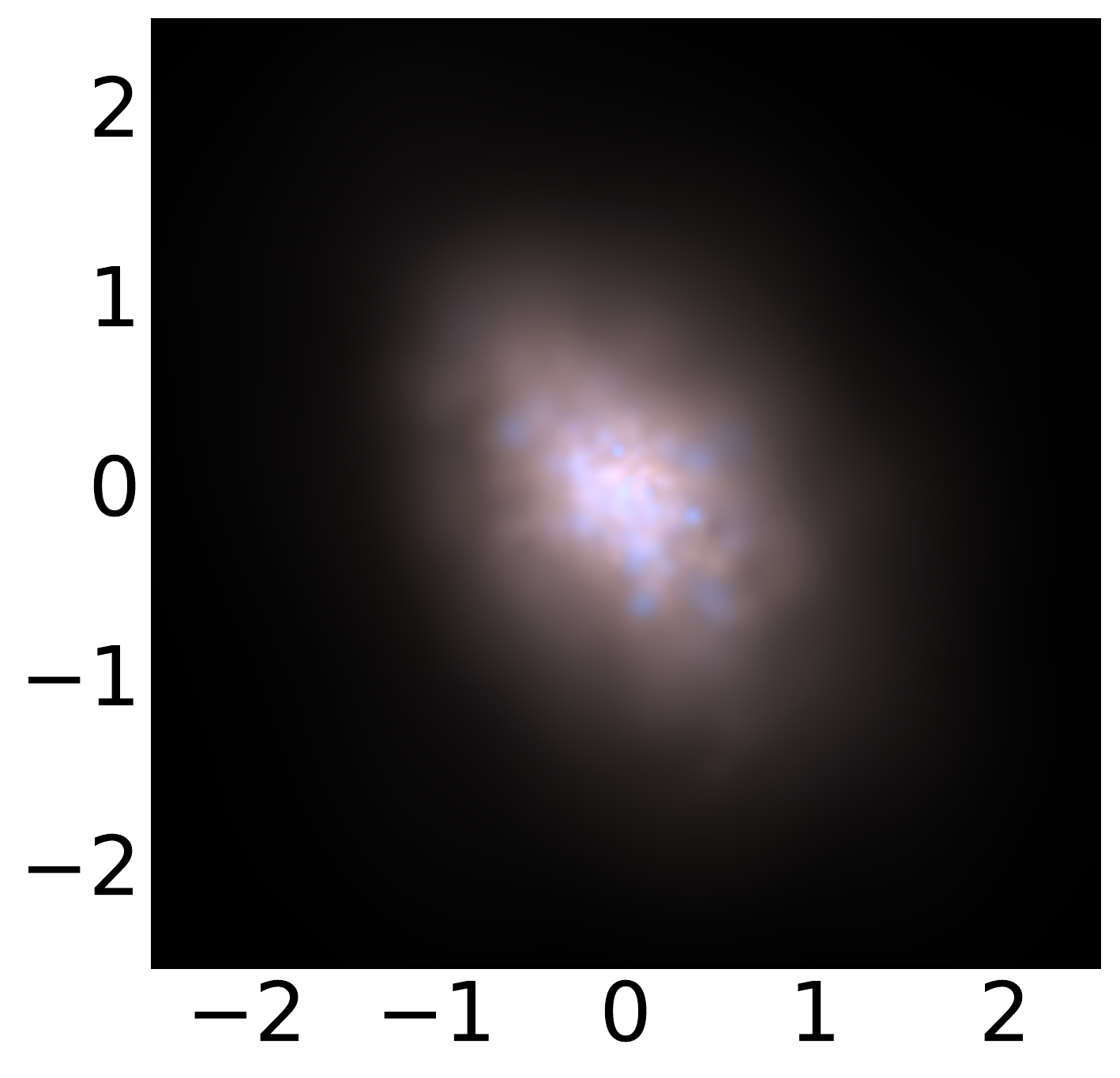}%
    \includegraphics[width=0.15\textwidth]{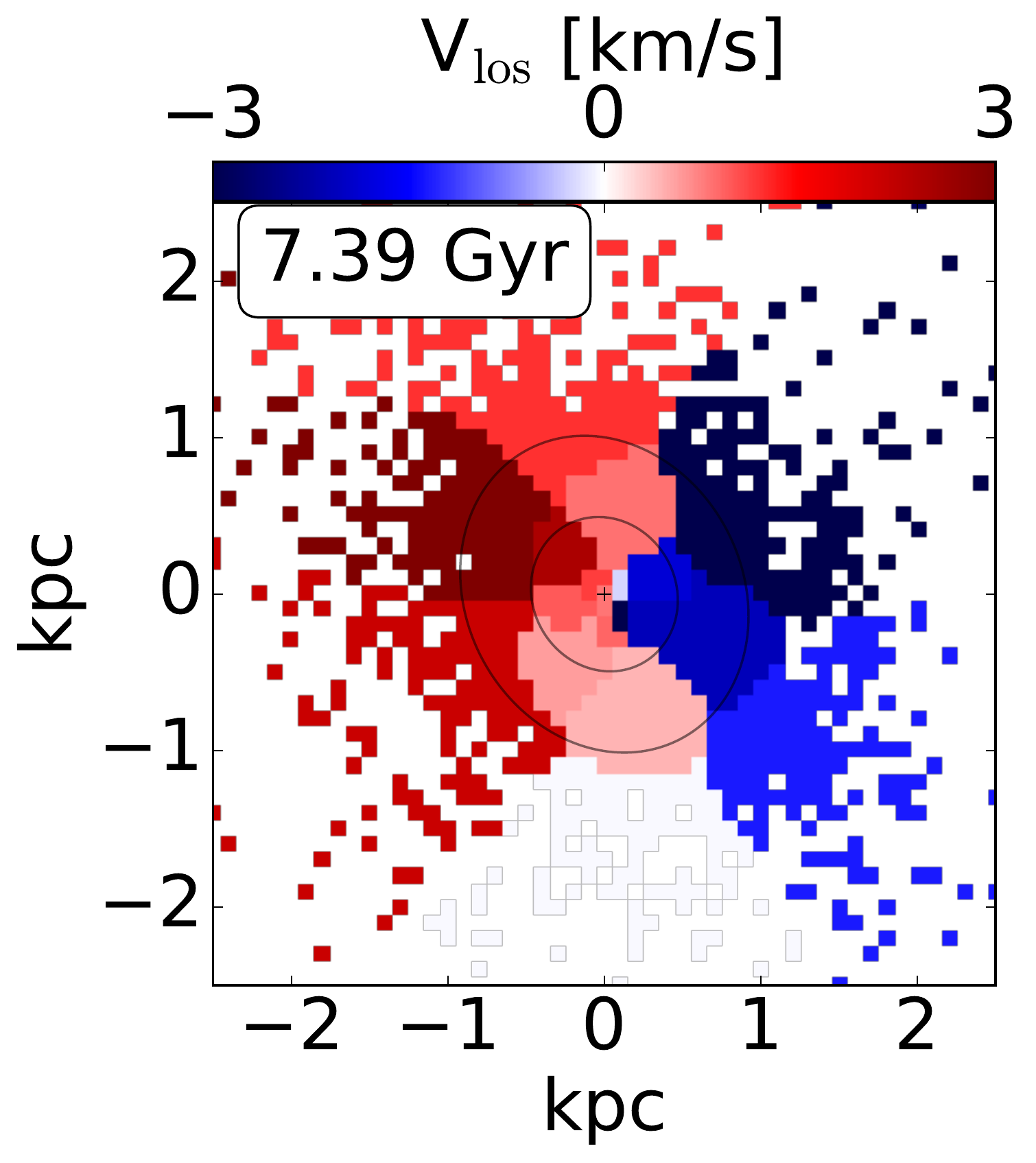}%
    \includegraphics[width=0.15\textwidth]{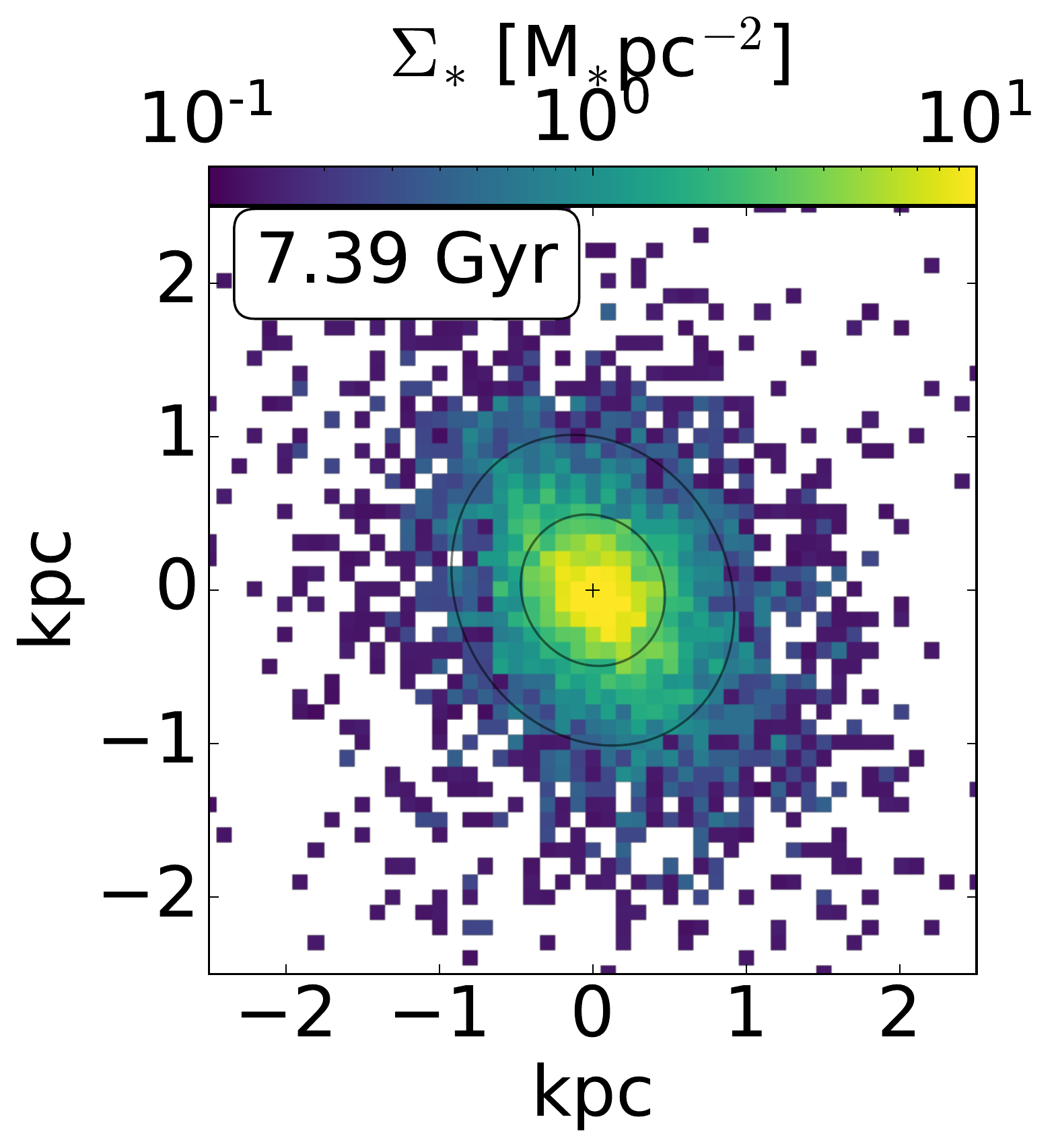}
    \includegraphics[width=0.15\textwidth]{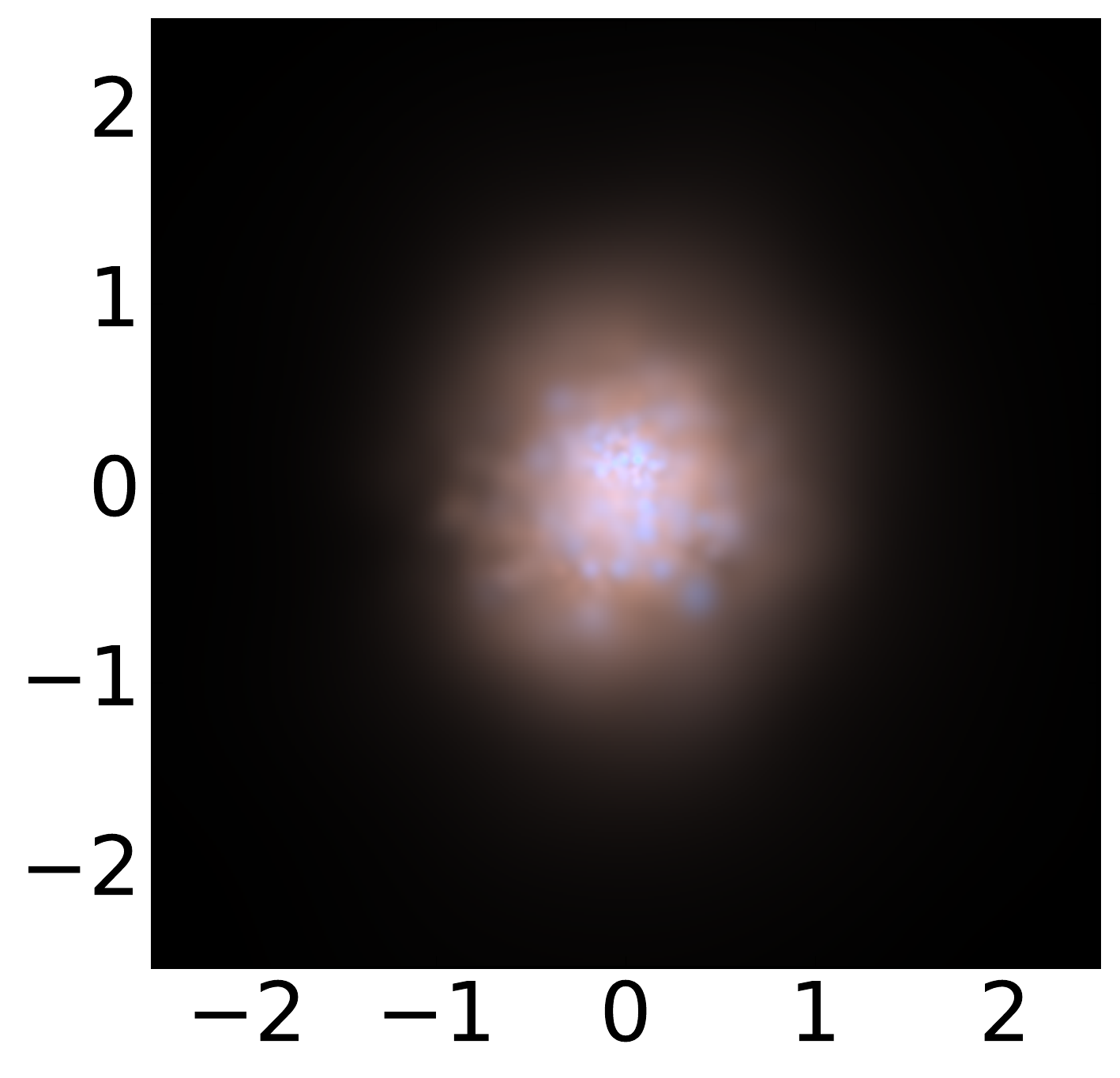}%
    \includegraphics[width=0.15\textwidth]{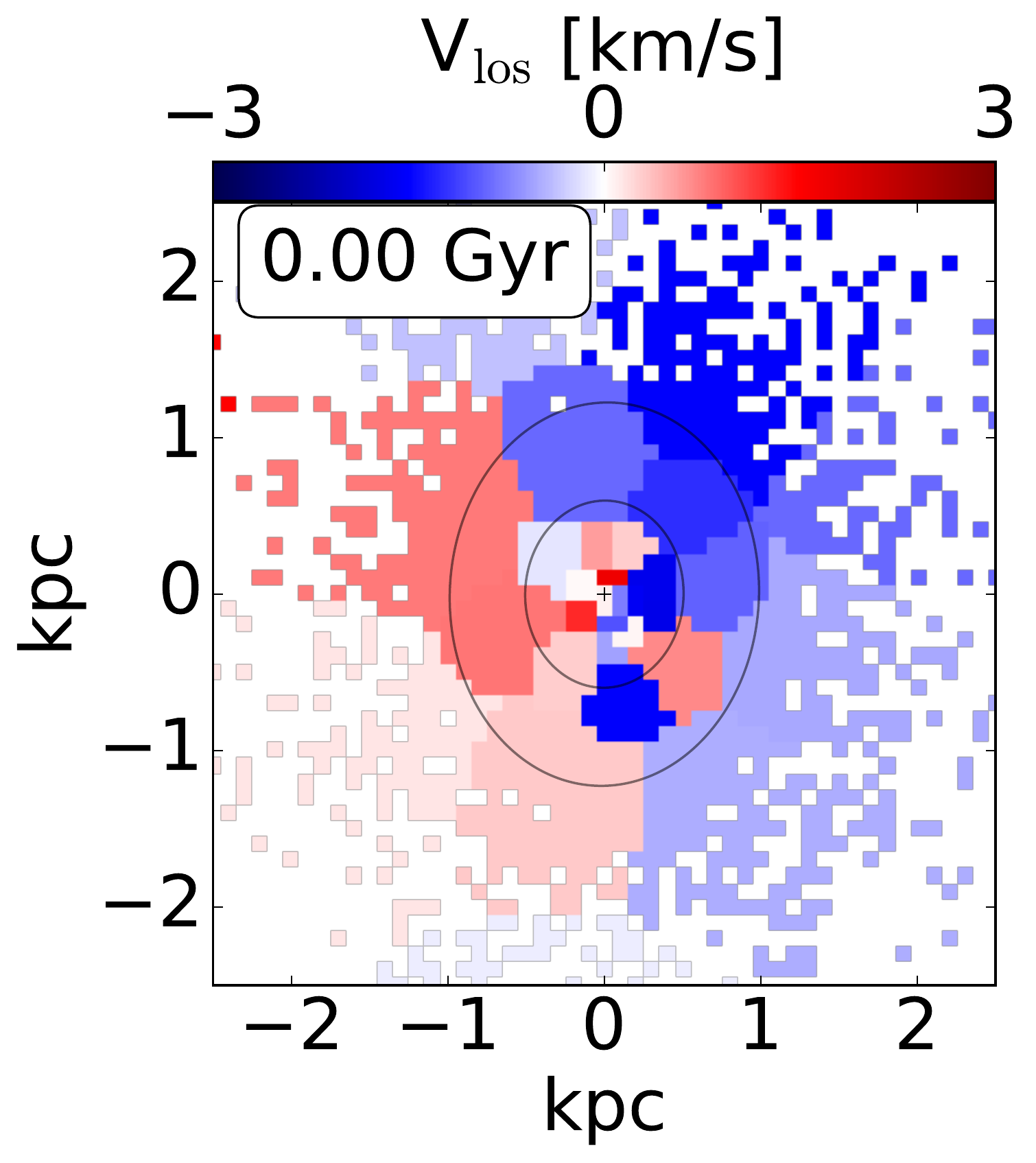}%
    \includegraphics[width=0.15\textwidth]{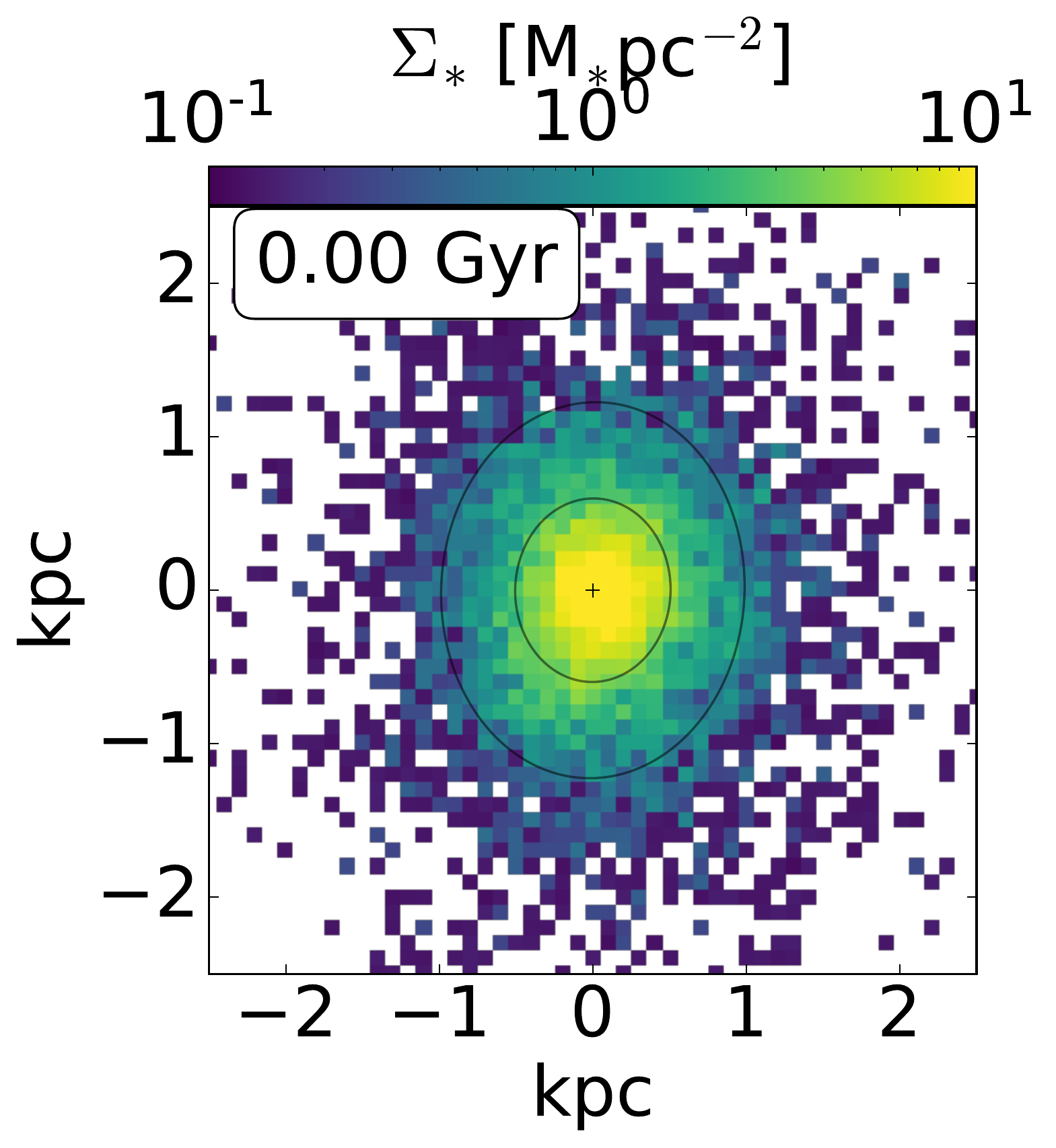}
    \caption{Properties of \texttt{h048} projected onto the line-of-sight. From left to right: RGB rendering of the stellar particles combining the $I$, $V$ and $B$ filters following the procedure described in \citealt{Lupton2004}; line of sight stellar velocity field; projected stellar density. Top: The main progenitor of the galaxy at $9.58$~Gyrs (before the merger). Middle: The main progenitor of the galaxy at $7.39$~Gyrs, at this time the main progenitor and the satellite are fully merged.  Bottom: galaxy at redshift $0$. The line of sight has been chosen to be perpendicular to the stellar angular momentum, which is aligned with the Y-axis in the projected maps. The stellar particles are distributed in a grid with $0.1$ kpc-wide bins. The velocity field has been constructed using the Voronoi tessellation algorithm of \citealt{Cappellari2003} ensuring a minimum number of $200$ star particles in each Voronoi bin. The velocity of each bin corresponds to the mass weighted average of the velocities of the stars belonging to the bin. The ellipses in the velocity and density maps indicate $1$ and $2$ half mass radius.}
    \label{fig:maps}
\end{figure}

\subsection{Significance}

Due to the low amplitude of the velocity gradient in the $3$D rotation curve, we are interested in proving the null hypothesis that the measured velocity gradient (i.e. $0.88\pm0.12$ km~s$^{-1}$~kpc$^{-1}$) can arise from a non-rotating distribution. 

To this aim, we have created $50000$ mock datasets with the same spatial coordinates of \texttt{h048} stellar particles, assigning a $3$D velocity to each "star"; each of the components of the velocity vector was sampled from Gaussian distributions with $0$~km~s$^{-1}$ mean and the standard deviation measured in \texttt{h048} along the $3$ principal axes.

We have measured the average rotation velocity gradient around the major axis  inside a sphere of $4$~kpc radius in each of the mock samples and found that none of them shows  a rotation velocity gradient as large as \texttt{h048} and only $22$ of them lie in the $3\sigma$ error interval. 

We have also compared the values of $(L_{\rm Major}/L_{\rm Tot})^{2}$ of \texttt{h048} with the one of the \textit{mocks}. We have chosen two apertures in which to measure this quantity: inside the $3$D half mass radius ($0.70$~kpc), and between the half mass radius and $4$~kpc.  The obtained values for \texttt{h048} are $0.769$ and $0.751$ for the two regions respectively. Only $0.9\%$ of the \textit{mocks} have a ratio $(L_{\rm Major}/L_{\rm Tot})^{2}$ as large as \texttt{h048}, or larger, in both regions at the same time. We conclude then that the signal detected has a low probability to appear in a non rotating system.

\subsection{Origin}

In order to investigate the causal link between the appearance of the prolate rotation and the main events in the life of  galaxy \texttt{h048}, we explore its mass accretion history. 
We constructed halo catalogs for all the snapshots in the simulation using the AHF Amiga Halo Finder \citep{Knollmann2009, Gill2004}, and merger trees using the freely available code MetroC++\footnote{https://github.com/EdoardoCarlesi/MetroCPP.git}. In the bottom panel of Fig.~\ref{fig:PA} we show the orbits of  several subhaloes that merged with the main progenitor of the galaxy  (note that only halos with more than $500$ particles are considered for the construction of the merger trees).  We also show the star formation history (SFH) of the main galaxy and the growth of its dark matter halo by means of its virial radius. It can be clearly seen that most of the mass assembly of \texttt{h048} occurs in the firsts Gyrs, having acquired half of its final, redshift $0$ mass about $10.5$ Gyrs ago. In this timeframe,  the galaxy undergoes several mergers and  its SF  is at its peak,  due to  the infalling material as well as to  the available star forming gas.

After the first $\sim2$ Gyrs,  the number of mergers is reduced, and the galaxy settles up in a triaxial configuration, with a mild rotation about the 3D minor axis reaching an amplitude of $3$ km s$^{-1}$ at $1$ kpc (top panels of Fig. \ref{fig:maps}, see also central column of Tab.~\ref{tab:dat} for a summary of the physical properties of the main progenitor). 

Around  $9.5$~Gyrs ago, the galaxy undergoes a massive accretion event with a luminous sub-halo with a stellar mass of $2.4\times10^5$\Msun and a halo mass of $3.4\times10^8$\Msun, with a mass ratio of 2:10 in  dark matter mass (see Tab.~\ref{tab:dat}). Fig.~\ref{fig:maps} shows the projected velocity field and stellar density before the merger (top), after the merger (middle) and at redshift $0$ (bottom). In these images the dynamical transformation from an oblate to a prolate rotator can be seen, coinciding exactly with the time of such major merger event.

This accretion event does not seem to have a tangible consequence on the SFH of the host.
However, there appears to be a clear correlation between the approach of the satellite and the change in the orientation of the angular momentum vector with respect to the direction of the principal axes, with the largest variations seen in correspondence to the pericentric passages. We therefore concluded that this event appears to be the responsible of the prolate rotation of  galaxy \texttt{h048}.

\section{Metallicity gradient}

\citet{Kacharov2017} point out that And~II and Phoenix, the only LG dwarf galaxies where prolate rotation has been detected, both share a similar steep metallicity gradient, which also appears to be the steepest gradient measured in LG dwarfs. One could then ask whether this is only a consequence of the pronounced age gradient existing in both systems  \citep[see e.g. ][]{delPino2017,Hidalgo2009}  or whether the putative massive accretion event responsible for the prolate rotation might have played a role. In order to investigate this aspect, we look at the properties of \texttt{h048}. 

\begin{figure}
    \centering
    \includegraphics[width=.9\columnwidth]{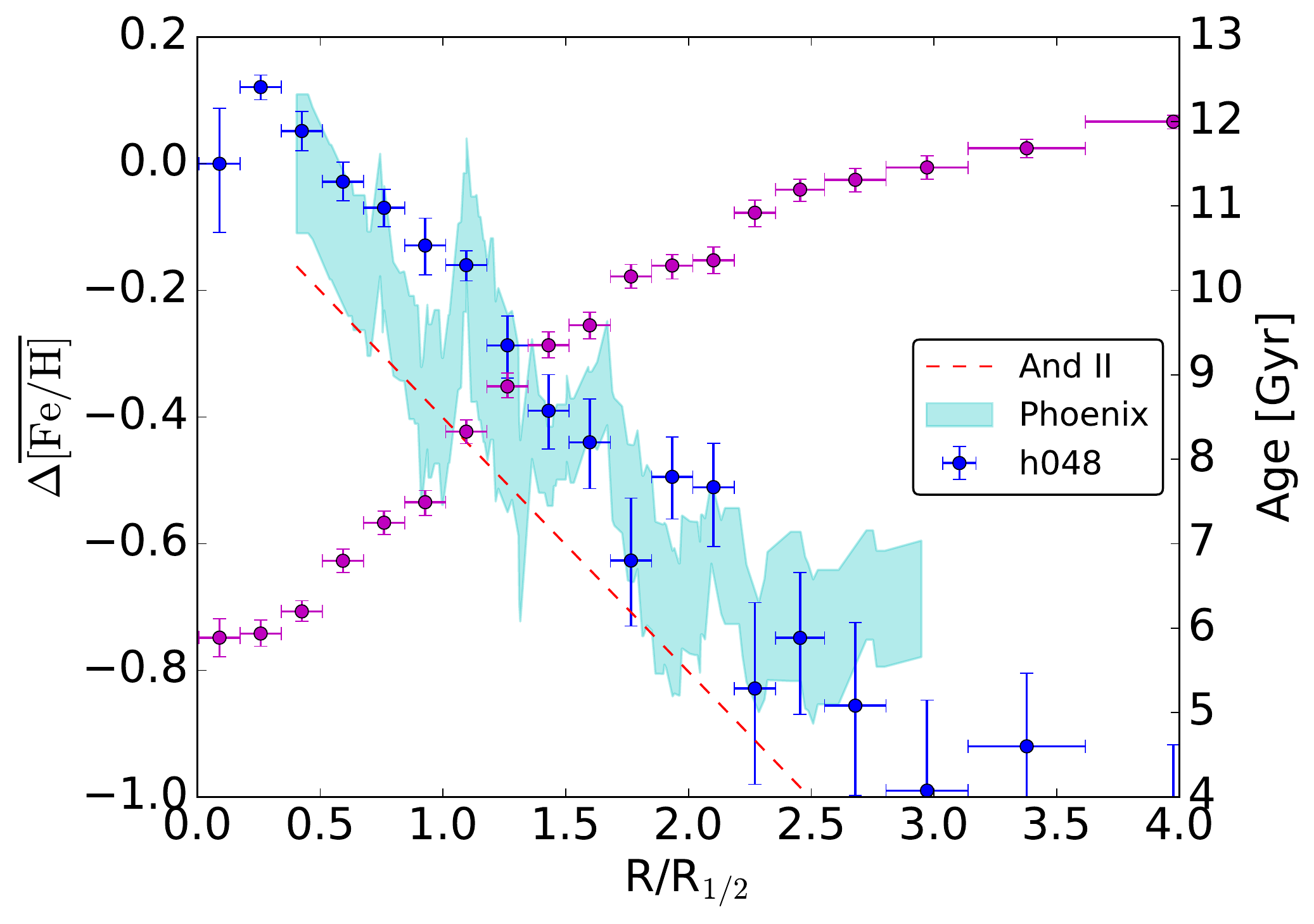} %
    \includegraphics[width=.85\columnwidth]{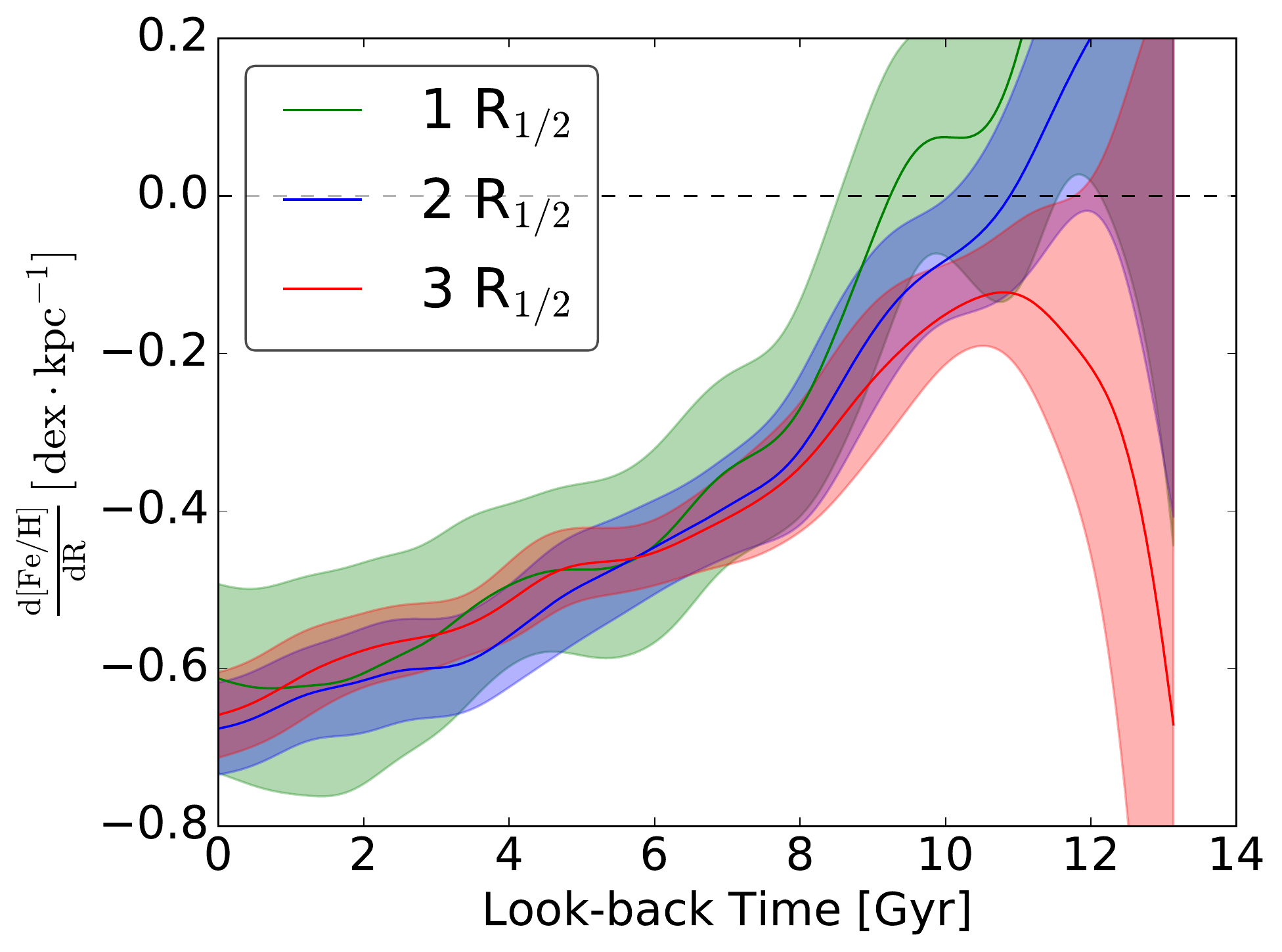}
    \caption{Top panel:  Average age (magenta) and metallicity (blue) projected radial profiles of \texttt{h048}, with the latter normalized to its central value. The error-bars correspond to the standard error of the average. The radial coordinate has been normalized to the half mass radius in order to compare with the  metallicity profiles of the Phoenix \citep[][]{Kacharov2017} and Andromeda~II \citep[][]{Ho2015} galaxies, in light-blue and dashed red respectively.   
    Bottom panel: the evolution of the metallicity gradient of the simulated galaxy as a function of look-back time, computed within $1$, $2$ and $3$ effective radii (green, blue and red respectively). The profiles have been smoothed via a Gaussian with standard deviation of $0.5$ Gyrs. The shaded regions correspond to the $1\sigma$ uncertainty over the smoothed evolution. 
   }
    \label{fig:Met}
\end{figure}

In the top panel of Fig.~\ref{fig:Met} we show the average metallicity and age profiles of the galaxy at $z=0$. The profiles are shown in projection, with the line-of-sight perpendicular to the stellar angular momentum vector, for a better comparison with the observations. The bins have been constructed in order to contain at least $200$ particles and have a minimum width of $0.1$ kpc in elliptical radius.

The simulated galaxy presents a strong metallicity gradient ($-0.67\pm0.03$~dex~kpc$^{-1}$ measured within an aperture of $2$ effective radii), which happens to be very similar to those of And~II and Phoenix, and the metallicity gradient clearly anti-correlates with the age profile. 

We then looked at the time evolution of the slope of the metallicity gradient (bottom panel of Fig.~\ref{fig:Met}).  It is apparent that, by the time the peak of SF has ceased, there is no metallicity gradient or only a very mild one (depending on the aperture in which it is calculated). Around the time of the main accretion event, a clear gradient is imprinted. After the merger, the gradient keeps evolving and increasing, smoothly. While the value of the gradient is somewhat dependent on the aperture, the main evolution is robust to it.
 We find that the merger dynamically heats the pre-existing population of stars, which move to wider orbits  \citep[compatible with the "outside-in" formation scenario of][]{Benitez-Llambay2016}. The residual SF, forms a population of younger metal rich stars that are no longer spatially mixed with the old population. This duality in the stellar population is able to generate a metallicity gradient of $\sim0.4$~dex~kpc$^{-1}$ in $\sim1$~Gyr.

After the merger the galaxy is not completely quenched (see bottom panel of Fig.~\ref{fig:PA}). We verified that the residual SF is limited to the very inner regions of the galaxy. This particular shape of the SF distribution is imprinted in the evolution of the metallicity gradient and is the responsible for the smooth steepening of the metallicity profile that happens after the merger. This process is in agreement with the correlation between the SFH and the metallicity gradient found in \citetalias{Revaz2018a}.
We conclude that while the accretion event responsible for the prolate rotation is not the only cause of the presence of a metallicity gradient, it  has played a significant role in steepening it, and that it is possible that the same  occurred to And~II and Phoenix.

\section{Discussion and Summary}
We performed a systematic search for prolate stellar rotation among $27$ simulated dwarf galaxies from the cosmological hydrodynamical zoom-in simulations described in \citetalias{Revaz2018a}. 
Those systems cover the stellar mass range between $3\times10^5$ and $5\times10^8$~\Msun, similar to LG dwarf galaxies. 
Out of those $27$ galaxies we find one prolate rotator, 
with a total mass of $2.8\times10^9$~\Msun and stellar mass of $1.4\times10^7$~\Msun at $z=0$. Even though the detected velocity gradient is of low amplitude, the analysis of mock datasets  suggests that it is unlikely that a velocity gradient of the same or larger amplitude would have arisen from a non-rotating system. 
To the best of our knowledge this is the first time that a prolate rotating galaxy with such a small mass  has been identified using non-idealized cosmological simulations.

The comparison between the merger history of \texttt{h048} and the evolution of the alignment between the principal axes of the stellar body with the direction of its  angular momentum vector clearly indicates that a major merger (mass ratio 2:10) is responsible for altering the internal kinematic properties of the host galaxy, changing it from an oblate to a prolate rotator. The  transformation is completed around $6$~Gyr ago, and the new configuration remains stable after that.  

This extends to a lower mass range the conclusions by \citetalias{Ebrova2017}.

The "host" and main satellite before the merger have different dark matter and stellar mass, and their spheroidal-like structure and low rotational vs pressure support ($V_{\rm rot}^{\rm max}/\sigma_{3D}$ $\sim 0.30$ for both the main progenitor and the satellite) resemble those of LG dwarf galaxies of similar stellar mass   \citep[see][]{Wheeler2017}.  This appears to be a more likely and realistic configuration than the highly flattened and rotating equal-mass stellar discs explored by \cite{Lokas2014} and \cite{Ebrova2015} to model And~II.

However, we point out that the level of rotation in \texttt{h048} at $z=0$ is much lower than what detected in And~II or Phoenix, the only known examples of prolate rotating dwarfs in the LG. In their study of prolate rotation on massive systems (M$_{\ast}>6\times10^9$~\Msun) from the ILLUSTRIS simulations, 
\citetalias{Ebrova2017}, show that it is not trivial to predict what configuration in the accretion/merger event would lead to the occurrence of prolate rotation and to what level of rotation.
Our goal was to explore whether prolate rotation could be detected at this mass scale from the accretions naturally occurring in cosmological simulations, and not to reproduce exactly the properties of And~II or Phoenix. In fact, environmental effects such ram-pressure stripping are essential to explain the detailed evolutionary path of And~II \citep[see][]{Fouquet2017}, while our simulated galaxies are isolated.
It remains to be explored what configurations,  arising in a $\Lambda$CDM Universe, could explain the fairly high rotation seen in And~II and Phoenix, and the likelihood of them happening.

Finally, the metallicity gradient in the simulated galaxy showing prolate rotation happens to be as steep as that measured for the Phoenix and And~II. 
The formation of this steep metallicity gradient appears to be the combined result of: the merger event responsible of the prolate rotation, which pushes to larger orbits the metal poor stars that were formed before $9.5$~Gyrs, and the subsequent SF, which is limited to the inner half mass radius and concentrated mainly in the innermost regions.

In this work, we showed that major mergers in dwarf galaxies offer an explanation for the formation of peculiar systems such as And II and Phoenix, characterized by steep metallicity gradients and prolate rotation.

\vspace{-.5cm}

\section*{Acknowledgements}
 SCB is supported by a FPI studentship from the Spanish Ministerio de Ciencia Innovación y Universidades (MCIU; PRE2018-086849). GB acknowledges financial support through the grant (AEI/FEDER, UE) AYA2017-89076-P, as well as by the Ministerio de Ciencia, Innovación y Universidades (MCIU), through the State Budget and by the Consejería de Economía, Industria, Comercio y Conocimiento of the Canary Islands Autonomous Community, through the Regional Budget. ADC is supported by a Junior Leader fellowship from `La Caixa' Foundation (ID 100010434), fellowship code  LCF/BQ/PR20/11770010.
\vspace{-1.cm}
\section*{Data Availability}
The data underlying this article will be shared on reasonable request to the corresponding author.
\bibliographystyle{mnras}
\bibliography{ProalteRotation}






\bsp	
\label{lastpage}
\end{document}